\documentclass[aps,pra,twocolumn,showpacs,amsmath,amssymb,longbibliography,10pt]{revtex4-2} 

\usepackage{lineno}
\usepackage{tgtermes}
\usepackage[utf8]{inputenc}
\usepackage{bm}
\usepackage{color}
\usepackage{makeidx}
\usepackage{amsfonts}
\usepackage{amssymb}
\usepackage{amsthm}
\usepackage{graphicx}
\usepackage{epstopdf}
\usepackage{epsfig}
\usepackage{subfigure}
\usepackage{amsmath,amssymb}
\usepackage[colorlinks=true,citecolor=blue,urlcolor=blue,linkcolor=blue]{hyperref}
\usepackage{xcolor}
\usepackage{subfigure}
\usepackage{framed}
\usepackage{booktabs}
\usepackage[normalem]{ulem}
\usepackage{bbold}
\usepackage{multirow}
\usepackage{pifont}

\newcommand{\bea}{\begin{eqnarray}}
\newcommand{\ea}{\end{eqnarray}}
\newcommand{\eea}{\end{eqnarray}}

\begin{document}

\title{Probing the Crossover between Dynamical Phases with Local Correlations in a Rydberg Atom Array
}

\author{Xiaofeng Wu}
\author{Xin Wang}
\author{Sixun Jia}
\author{Bo Xiong}
\email{boxiong@whut.edu.cn}
\affiliation{Department of Physics, Wuhan University of Technology, Wuhan 430070, China}

\date{\today}

\begin{abstract}
The experimental detection of non-equilibrium quantum criticality remains a challenge, as traditional signatures like dynamical quantum phase transitions rely on hard-to-measure global properties. Here, we demonstrate that local connected correlation functions provide a superior, practical means to directly probe the dynamics of magnetic order in a quenched Rydberg atom array. Using a Magnus expansion formalism, we derive analytic expressions for these correlations that capture a smooth crossover from antiferromagnetic to ferromagnetic dominance. Our analytic results, which reveal the critical parameter relationship $U_{c}(\delta)$, are validated against exact numerical simulations and exhibit robustness to finite-size effects. By shifting the focus from global singularities to local correlations, our protocol establishes a direct and feasible path to observe the rich critical dynamics in scalable quantum simulators.

\end{abstract}

\maketitle

\section{Introduction}


The study of out-of-equilibrium dynamics in strongly correlated systems bridges fundamental physics and cutting-edge technology \cite{Polkovnikov2011, 47, 48, 49, 50}. 
It uncovers new states of matter \cite{4, Huang}, reshapes out understanding of thermalization and quantum chaos \cite{40, DAlessio2016, Adam}, and drives innovations in quantum engineering and materials science \cite{Tauchert2022}. 
A central challenge in this field is to understand the emergence of spatial correlations and entanglement in systems driven far from equilibrium. Such dynamics often exhibit critical phenomena and abrupt transitions, revealing behaviors with no direct analogue in conventional equilibrium physics.

A cornerstone of these time-dependent critical phenomena is the concept of dynamical quantum phase transitions (DQPTs), which generalize the notion of phase transitions to the temporal domain. Conventionally, DQPTs are identified by non-analyticities in the dynamical free energy density, derived from the Loschmidt amplitude \cite{Wong},
\begin{equation}
    \mathcal{L}(t) = \langle \psi_{0} | e^{-iHt} | \psi_{0} \rangle,
\end{equation}
which measures the overlap between the time-evolved state and the initial state $|\psi_{0}\rangle$. Singularities in the associated rate function,
\begin{equation}
    \lambda(t) = - \lim_{N \to \infty} \frac{1}{N} \ln |\mathcal{L}(t)|^2,
\end{equation}
signal the occurrence of a DQPT at critical times $t_{c}$, often interpreted as moments where the evolved state becomes orthogonal to the initial state \cite{Heyl, Heyl_2018}. While this framework provides a powerful formal analogy to equilibrium thermodynamics, its practical application faces significant challenges. The quantity $\lambda(t)$ is a global measure, and its divergences correspond to idealized, singular transitions—such as a complete inversion from a perfect ferromagnet to a perfect antiferromagnet.

In realistic quantum simulators, however, the dynamics are often richer and more nuanced. The time-evolved state is typically a complex superposition, and the system may undergo a smooth \emph{crossover} where the dominant character of its correlations evolves gradually—for instance, from predominantly antiferromagnetic (AF) to ferromagnetic (FM) order—without a sharp global singularity \cite{king}. This continuous evolution of magnetic order, dictated by the underlying quench parameters, represents a fundamental aspect of non-equilibrium physics that is not fully captured by the Loschmidt paradigm. Probing these dynamical crossovers demands tools that are simultaneously sensitive to local order, experimentally feasible, and capable of providing analytical insight into the critical parameter dependencies governing the transition.

In this work, we bridge this gap by investigating the crossover between dynamical magnetic phases in a Rydberg atom quantum simulator. We demonstrate that \emph{local connected correlation functions} serve as superior, practical probes for tracking the evolution of magnetic order during a parametrically driven quench from a paramagnetic to an AF regime. Using a \emph{Magnus expansion approach}, we derive closed-form analytic expressions for these correlations, which not only show quantitative agreement with exact numerical simulations but also directly reveal the functional relationship $U_{c}(\delta)$ between critical interaction strength and detuning at which the AF-to-FM crossover occurs. Our results establish a robust framework for detecting and characterizing dynamical phase behavior in finite quantum systems, offering a directly applicable protocol for current experimental platforms where global measurements like the Loschmidt echo remain challenging.

This paper is structured as follows. In Sec. \ref{sec2}, we begin by introducing the Ising-type model Hamiltonian implemented in a Rydberg atom system, along with the application of the Magnus expansion (ME) to characterize the system's dynamical evolution via connected correlation functions. In Sec. \ref{sec3}, we analyze the crossover between different dynamical phases during the non-equilibrium evolution of the system, as captured by the nearest-neighbor connected correlation function, from both analytical and numerical perspectives. This behavior is compared with results obtained from longer-range correlations, and the effects of larger system sizes and different lattice geometries on the connected correlation function are also discussed. Finally, Sec. \ref{sec4} summarizes the conclusions of our study.

\section{Model and method}

\label{sec2}

\subsection{Ising-like Rydberg system}

Our model builds upon recent advances in controlling individual $^{87}\text{Rb}$ atoms in programmable two-dimensional optical tweezer arrays, with each tweezer confining a single atom. This setup offers a highly controllable environment for probing nonequilibrium quantum dynamics. The atoms are initially prepared in the electronic ground state $|0\rangle$ via optical pumping and then coherently coupled to a Rydberg state $|1\rangle$ through a two-photon transition with Rabi frequency $\Omega$ and detuning $\delta$.
The system is described by the effective Hamiltonian:
\begin{equation}
\label{3}
H = \frac{\hbar\Omega(t)}{2} \sum_{i} \sigma_{i}^{x} - \hbar\delta(t)\sum_{i}n_{i} + U\sum_{\langle ij \rangle}n_{i}n_{j},
\end{equation}
where $\sigma_{i}^{x} = |0\rangle\langle 1|_{i} + |1\rangle\langle 0|_{i}$ is the $x$-Pauli operator, driving coherent transitions between $|0\rangle$ and $|1\rangle$ at site $i$, and $n_{i} = |1\rangle\langle 1|_{i} = \frac{1}{2}(1 - \sigma_{i}^{z})$ is the Rydberg projection operator. The interaction term $U_{ij} \sim 1 / r_{ij}^6$ captures the van der Waals interaction between Rydberg excitations at sites $i$ and $j$, with $r_{ij}$ being the interatomic distance.
A key consequence of these interactions is the Rydberg blockade effect: when atoms lie within a critical distance $R_b \approxeq r_{\langle ij \rangle}$, their strong interaction suppresses simultaneous Rydberg excitation, effectively enforcing a nearest-neighbor exclusion constraint. This mechanism facilitates the emergence of correlated quantum phases and enables quantum simulation of strongly interacting spin systems.
Using this platform, a variety of interaction-driven many-body phenomena have been observed, including DQPTs \cite{Heyl, Heyl_2018, PhysRevLett.115.140602}, quantum many-body scars \cite{4}, and quantum phase transitions \cite{Guardado2018, Lienhard2018, king}. In this framework, the Hamiltonian maps onto a quantum Ising model with a transverse field $B_\perp \propto \Omega$, a longitudinal field $B_\parallel \propto \delta$, and spin-spin couplings $U$. 

\subsection{Magnus expansion}

\begin{figure*}[htpb]
	\centering
	\includegraphics[scale=1.0]{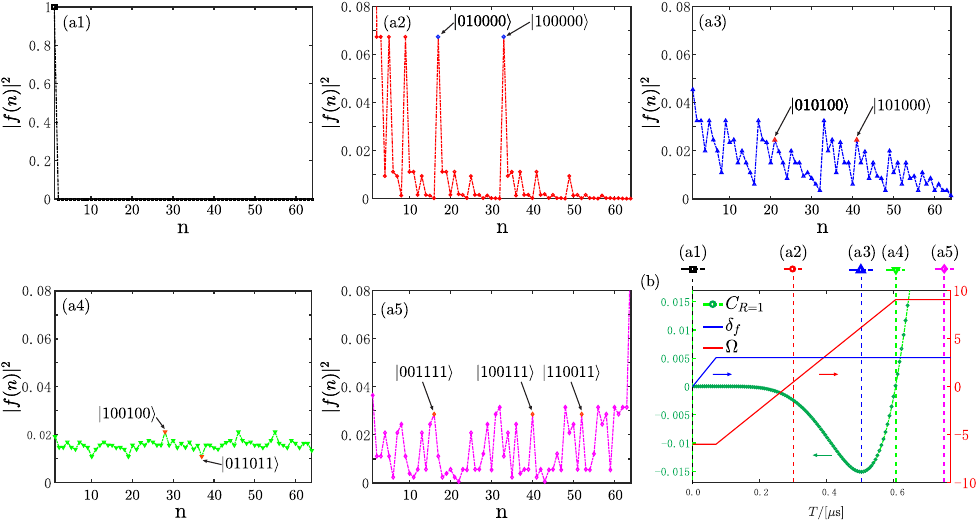}
	\caption{Temporal density distribution $|f(n)|^{2}$ for 6 sites with $U/h=3 \, \rm MHz$ at $t=0 \, \mu s$ (a1), $t=0.3 \, \mu s$ (a2), $t=0.5 \, \mu s$ (a3), $t=0.6 \, \mu s$ (a4), $t=0.75 \, \mu s$ (a5). (b) The quench protocol, showing the time-dependent parameters $\delta(t)$ and $\Omega(t)$ (right vertical axis), and the resulting dynamics of the nearest-neighbor correlation function $C_{R=1}$ (left vertical axis).}
    \label{Fig1}
\end{figure*}

The nonequilibrium dynamics of strongly correlated many-body systems presents significant theoretical challenges, primarily due to the exponential growth of the Hilbert space with system size. Notwithstanding this complexity, exact analytical solutions have been constructed for systems with local interactions and finite on-site Hilbert spaces on n-dimensional hypercubic lattices \cite{45}. Concurrently, the ME has emerged as a powerful analytical tool for tackling such dynamics. While its variant, the Floquet-Magnus expansion, is tailored for periodically driven systems \cite{KUWAHARA201696,RODRIGUEZVEGA}, the standard ME framework has proven highly effective in describing quantum quenches. Recent advances demonstrate that this ME approach can be employed to derive analytical expressions for key dynamical observables--particularly the connected spin-spin correlation function--in Ising-like models, achieving results in excellent agreement with exact numerical solutions \cite{10}. Within this present ME framework (or present theoretical framework), it has also been established that the AF correlation magnitude at a fixed Manhattan distance obeys a universal superposition principle, which remains robust against variations in path equivalence, lattice geometries, and quench protocols \cite{Xinwang}.

Building upon these developments, we apply ME to explore the crossover between dynamical magnetic phases in an Ising-like model realized by a Rydberg atomic system. As a methodological foundation, ME provides a systematic approach for constructing approximate exponential representations of the time-evolution operator for systems governed by time-dependent Hamiltonians \cite{2}. This technique has found widespread application across diverse fields, from atomic and molecular physics \cite{20,21} to nuclear magnetic resonance \cite{22} to quantum electrodynamics and elementary particle physics \cite{23,24}.

Under the condition $\int_{0}^{T} ||H(t)||_{2} \, dt < \pi $, where $||H||_{2}$ denotes the euclidean/spectral norm of $H$ defined as the squared root of the largest eigenvalue of positive semi-definite operator $H^{\dagger}H$, the ME treatment of the many-body propagator yields the time-evolution operator $\hat{U}(T)={\rm exp} [-iT\hat{H}(T)]$. Here, $\hat{H}(T)$ is given by a series expansion involving nested commutators of the time-dependent Hamiltonian (\ref{3}), i.e., $\bar{H}(T)=\sum_{k=1}^{\infty}\bar{H}_{k}(T)$. To maintain computational tractability, we truncate the ME at second order, with $\bar{H}_{1}=\frac{1}{T}\int_{0}^{T}H(t_{1})dt_{1}$ and $\bar{H}_{2}=\frac{i}{2T}\int_{0}^{T}[\int_{0}^{t_{1}}H(t_{2})dt_{2},H(t_{1})]dt_{1}$, where the commutator is defined as $[\hat{A},\hat{B}]=\hat{A}\hat{B}-\hat{B}\hat{A}$. A recent work has examined the necessary of a second-order ME for a frequency chirped quantum two-level system, comparing it with the rotating wave approximation and exact solution \cite{Begzjav}.

To model experimentally relevant scenarios, we consider a linearly ramped detuning $\delta(t)=\frac{\delta_{f}-\delta_{0}}{T}t+\delta_{0}$, where $\delta_{0}$  and $\delta_{f}$ denote the initial and final detuning values, respectively. One can then readily derive (with $\hbar = 1$), $\bar H_{1} = \frac{\Omega}{2}\sum_{i}\sigma_{i}^{x} - \delta_{\rm avg}\sum_{i}n_{i}+\sum_{\langle ij\rangle}U_{ij}n_{i}n_{j}$ and $\bar H_{2} = \frac{\Omega}{24}(\delta_{f}-\delta_{0})T\sum_{i}\sigma_{i}^{y}$ with $\delta_{\rm avg}=\frac{\delta_{0}+\delta_{f}}{2}$. Note that the final term in $\bar H$ depends linearly on $T$, stemming from the second-order ME. Furthermore, by expanding the matrix exponential as $\hat U(T) = \sum_{n=1}^{\infty}\frac{(-iT)^{n}}{n!}\bar H^{n}(T)$, we observe that higher-order terms in $T$ emerge, which will contribute to the correlation function.

In contrast to the standard Ising model, where AF correlations can be directly characterized by local density expectations such as $\langle n_{1}n_{2} \rangle$, the situation in our driven system is more subtle. Due to the presence of nontrivial magnetic fields and time-dependent driving—encoded in the effective Hamiltonian $\bar{H}(T)$ obtained from the ME—a more refined diagnostic is required to properly capture ordering. In this context, AF correlations are quantified through the connected spin-spin correlation function:

\begin{equation}
C({\bf r})=\frac{1}{N_{{\bf r}}} \sum_{( ij ), |{\bf r}_i-{\bf r}_j|=r} [\langle n_{i}n_{j} \rangle - \langle n_{i} \rangle \langle n_{j} \rangle],
\end{equation}
where ${\bf r}=(ka,la)$ denotes the displacement vector, $N_{\bf{r}}$ counts atom pairs separated by $\bf{r}$ and the summation runs over all such pairs in the lattice.

To characterize the dynamics governed by the critical parameters in strongly driven systems, we focus on the connected spin-spin correlation function, which captures nontrivial spatiotemporal ordering beyond local observables. Based on the effective Hamiltonian, we carry out systematic analytical calculations for various lattice geometries and obtain universal expressions for the AF correlation function $C_{R}$. Explicit results are derived for any sites ($C_{R}$):
\begin{equation}
\centerline{\includegraphics[scale=1.0]{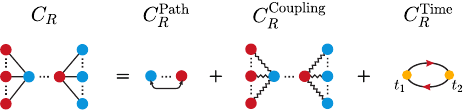}},\notag
\end{equation}
where the solid lines between two reference balls label the shortest paths and the wave lines denote the coupling from the nearest-neighbor sites to the shortest paths.
To clarify the physical origin of different contributions to the correlation function, we analyze $C_{R=1}$ within the ME framework. The correlation is decomposed into three components: (a) Path contribution ($C_{R=1}^{\rm Path}$): Dominated by the leading-order ME term, this captures propagation along the shortest path between adjacent lattice sites. (b) Coupling contribution ($C_{R=1}^{\rm Coupling}$): Arising from higher-power terms in the first-order ME, it quantifies environmental coupling effects -- specifically, perturbations from neighboring sites on the shortest paths. (c) Time-dependent contribution ($C_{R=1}^{\rm Time}$): Generated by the second-order ME term, this reflects non-commutative quantum dynamics due to the nonzero commutator $\left[ H(t_{1}), H(t_{2})\right] \ne 0$ for $t_{1} \ne t_{2}$. Physically, it encodes temporal interference from interaction time-ordering, leading to corrections in correlation buildup that are essential in nonequilibrium regimes. 
The distinct components of $C_{R}$ provide a clear diagnostic for tracking nonequilibrium critical behavior, allowing us to monitor the onset and evolution of the dynamical crossover through the buildup of correlations. Crucially, the observation of such a crossover requires a significant breakdown of adiabaticity. From our ME, we explicitly recover the standard adiabatic condition and confirm that our chosen quench protocol strongly violates it [see Appendix \ref{SM4}]. This established nonadiabaticity is essential, as it provides the necessary dynamical drive for the system to explore competing orders and exhibit the phase crossover studied in the following sections.

\section{Results and discussion}

\label{sec3}

\subsection{A typical dynamical process}

To uncover the microscopic mechanism driving the dynamical phase transition, we examine the evolution of the probability density distribution over the many-body Hilbert space. This analysis reveals how the system's population shifts among different quantum configurations.
We illustrate this process using a six-site lattice as a concrete example.
With each atom occupying either the ground or Rydberg state, the system possesses $2^{6}=64$ possible configurations, whose probability dynamics are central to the transition.

Starting with all atoms in the ground state $|\psi(t=0)\rangle=|000000\rangle$ [see Fig.\,\ref{Fig1}\,(a1)], we ramp linearly up the Rabi frequency $\Omega$ from zero to $\Omega_{\rm max}/(2\pi)= 3 \, \rm MHz$ over a duration $t_{\rm rise}=0.07 \, \rm \mu s$, maintaining a constant negative detuning $\delta_{0}/(2\pi)=- 6 \, \rm MHz$. Although $\Omega$ prefer the equal occupation in $|0\rangle$ and $|1\rangle$ states, the large negative detuning $\delta_{0}$ suppresses greatly the occupation in $|1\rangle$ so the nearest-neighbor correlation function $C_{R}\approx0$ in this stage [see Fig.\,\ref{Fig1}\,(b)]. At the sequent stage, we hold $\Omega$ at $\Omega_{\rm max}$ for a time interval $t_{\rm sweep}=0.53 \, \rm \mu s$ while linearly sweeping the detuning from $\delta_{0}$ to a final positive value $\delta_{f}/(2\pi)=9.0 \, \rm MHz$, thereby driving the system into the AF phase. The spin-spin correlation $C_{R=1}$ varies from $0$ to negative value, reaching the maximum AF at $t=0.5 \rm \mu s$ with $C_{R=1}=-0.015$. Correspondingly, the density distribution of the state is initially dominated by components with AF entanglement, such as the state $|100000\rangle+|010000\rangle$ [see Fig.\,\ref{Fig1}\,(a2)]. As the dynamics evolve, this AF character is progressively enhanced through the buildup of local $|01\rangle+|10\rangle$ superpositions on neighboring sites [see Appendix \ref{SM1}], leading to the participation of more extended AF entangled states such as $|101000\rangle+|010100\rangle$ [see Fig.\,\ref{Fig1}\,(a3)]. At the final stage of the quench, the parameters are held constant at $\Omega=\Omega_{\rm max}$ and $\delta=\delta_{\rm max}$ until $t=1 \, \rm \mu s$ to monitor the state evolution. In principle, a sufficiently slow (adiabatic) quench could prepare the system in the AF phase. However, the driven process here is nonadiabatic so that the system does not reach the AF phase but instead enters an excited FM region, evidenced by a positive correlation $C_{R=1}>0$. The density distribution evolves sequentially: starting from an AF-dominated pattern, it transitions through a PM regime ($C_{R=1} \approx 0$) where the many-body state approaches a nearly uniform superposition of all basis states [see Fig.\,\ref{Fig1}\,(a4)], and finally settles into an FM-dominated structure characterized by local FM superpositions such as $|00\rangle+|11\rangle$ [see Fig.\,\ref{Fig1}\,(a5)].

\begin{figure}[htpb]
	\centering
	\includegraphics[scale=1.0]{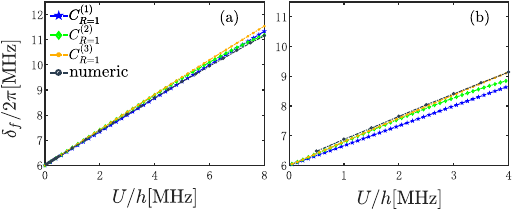}
	\caption{The crossover is mapped as a function of interaction and detuning, with case studies at $\Omega/(2\pi) = 1 \, \rm MHz$ (a) and $\Omega/(2\pi) = 2 \, \rm MHz$ (b) upon quench completion. Comparison between the analytical expansion terms $C_{R=1}^{(n)}$ (colored dotted lines) and numerical results (black dotted lines) for the nearest-neighbor correlation function.}
    \label{fig2}
\end{figure}



\subsection{Identifying crossover via local correlations}

\begin{figure}[htpb]
	\centering
	\includegraphics[scale=1.0]{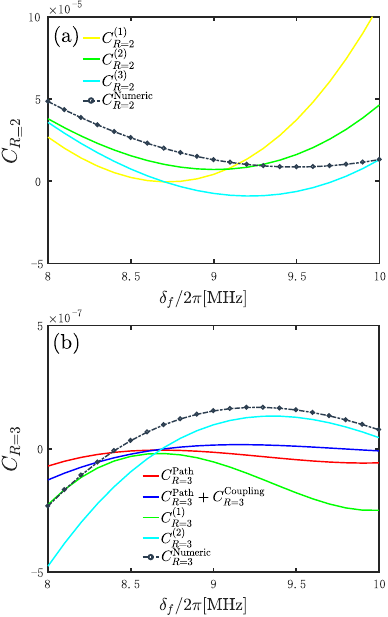}
	\caption{Evolution of the long-range correlation function with detuning at the end of the quench. The next-nearest-neighbor $C_{R=2}^{(n)}$ (a) and next-next-nearest-neighbor $C_{R=3}^{(n)}$ (b) correlation functions are investigated at the fixed interaction strength $U/h=3\,\rm MHz$ and Rabi frequency $\Omega/2\pi = 2\, \rm MHz$. The analytical results from the first-order ($\tilde{C}_{R}^{(n)}$) and second-order ($C_{R}^{(n)}$) ME are compared against numerical simulations.}
    \label{fig3}
\end{figure}

\begin{figure*}[htpb]
	\centering
	\includegraphics[scale=1.0]{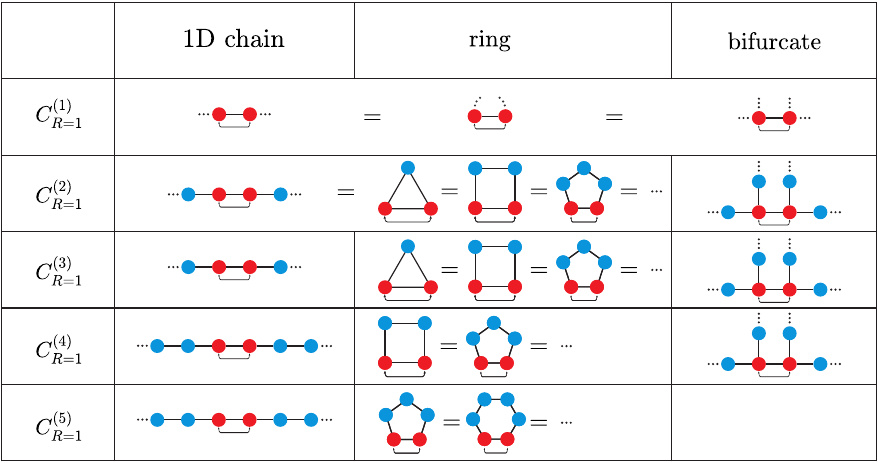}
	\caption{Analytic expressions for the $n$-th order approximation of the nearest-neighbor correlation function, $C_{R=1}^{(n)}$, are identical for the three lattice geometries shown (1D chain, ring, bifurcate), provided the system size $N$ meets a minimum threshold. Dots indicate the lattice can be extended arbitrarily. The required minimum size increases with the order $n$: for $C_{R=1}^{(1)}$, $N \geq 2$ suffices for all geometries; for $C_{R=1}^{(2)}$, $N \geq 4$ (chain), $N \geq 3$ (ring), and $N \geq 6$ (bifurcate).}
    \label{fig4}
\end{figure*}

The preceding analysis demonstrates that the smooth dynamical crossover from AF to FM ordering is jointly controlled by the detuning $\delta$ and the interaction strength $U$ for a fixed Rabi frequency $\Omega$. We find that, within a characteristic time window, the crossover occurs when $U$ reaches a specific value intrinsically tied to $\delta$. To elucidate the quantitative relationship between $U$ and $\delta$, we employ the ME to derive analytical expressions at various orders and compare them with exact numerical simulations. Our numerical results further confirm that, in the short-time regime relevant to the crossover, the evolution of the connected correlation function is independent of system size--exhibiting clear universal scaling--which validates the reliability of our finite-size simulations [see Appendix \ref{SM5}].

As shown in Fig.\,\ref{fig2}, the analytical results for the nearest-neighbor correlation function agree quantitatively with the numerical benchmarks. We define the $n$-th order approximation of the correlation function, denoted as $C_{R=1}^{(n)}$, as follows (see Appendix \ref{SM2} for details): 
(i) $n=1$ accounts for the direct nearest-neighbor interaction and the leading-order temporal interference due to interaction time-ordering, i.e., contributions labeled as $C_{R=1}^{\rm Path}$ and $C_{R=1}^{\rm Time}\left[O(T^{12})\right]$.
(ii) $n=2$ incorporate progressively higher-order couplings along the shortest paths between neighboring sites.
(iii) $n=3$ further includes perturbations from next-nearest neighbors along these paths.

For $\Omega/2\pi = 1 \, \rm MHz$ [Fig.\,\ref{fig2}\,(a)], the critical condition $C_{R=1}=0$ is satisfied when $U$ scales nearly linearly with $\delta$, approximately following $U\approx3\delta$ [see Appendix \ref{SM2}]. Although higher-order expansions do not significantly improve accuracy in this case, the situation changes markedly for $\Omega/(2\pi) =2\, \rm MHz$ [Fig.\,\ref{fig2}\,(b)]. Here, higher-order correlations become essential for accurately capturing the critical transition point, with results converging toward the exact numerical solution as $n$ increases. This trend indicates that larger Rabi frequencies enhance effective interactions among the shortest paths, making perturbative contributions from neighboring sites non-negligible.

We now extend our analysis to the next-nearest neighbor ($C_{R=2}$) and next-next-nearest neighbor ($C_{R=3}$) correlation functions, focusing on their evolution as $C_{R=1}$ transitions from negative (AF) to positive (FM) values. Unlike $C_{R=1}$, the analytic condition $C_{R=2}=0$ yields multiple solutions, making it unsuitable for uniquely identifying the transition point. Therefore, we adopt an alternative approach: for fixed parameters $U/h = 3\, \rm MHz$, $\Omega/(2\pi) = 2 \, \rm MHz$, and a sweep time $t_{\rm sweep}=0.53\mu s$, we employ the ME to compute $C_{R=2}$ and $C_{R=3}$ across the phase transition region identified by $C_{R=1}$.

As shown in Fig.\,\ref{fig3} (a), both analytical and numerical results show that $C_ {R=2}$ exhibits a non-monotonic dependence on the final detuning $\delta_{f}$, initially decreasing to a minimum before rising again as $\delta_{f}$ increases from $8$ to $10\, \rm MHz$. The ME successfully captures this overall trend. Moreover, the convergence behavior of the ME is physically revealing: higher-order expansions (larger $n$) achieve quantitative agreement with exact numerics over a wider parameter range. This convergence indicates that at larger detunings, the system's behavior is governed by a competition between different physical processes. Specifically, the contributions from neighboring sites beyond the shortest path--which are incorporated at higher orders--become increasingly significant, highlighting the essential role of these extended correlations in determining the system's non-equilibrium response.

In contrast, $C_{R=3}$ exhibits the opposite trend: it increases with $\delta_{f}$ before reaching amaximum and then decreasing [Fig.\,\ref{fig3}\,(b)]. The ME reveals why an accurate description is challenging: lower-order approximations--considering only the shortest path (red); or including temporal interference (green); or adding nearest-neighbor couplings along the shortest path (blue) -all fail to replicate the numerical results. Quantitative agreement is achieved only with $C_{R=3}^{(2)}$, which incorporates perturbations from next-nearest neighbors. This provides compelling evidence that correlations at this range are fundamentally governed by extended coupling pathways beyond the shortest link.

Notably, the extremal points of these correlation functions occur near the $C_{R=1}=0$ transition point: specifically, $C_{R=2}$ reaches its minimum at $\delta_{f}/(2\pi)\approx9.5\,\rm MHz$, and $C_{R=3}$ reaches its minimum at $\delta_{f}/(2\pi)\approx9.2\,\rm MHz$, while $C_{R=1}=0$ occurs at $\delta_{f}/(2\pi)\approx9.0\,\rm MHz$. This clustering of extrema indicates that during the AF-FM transition, all short-range correlation functions tend to approach their minimal values within a narrow parameter range. This regularity is found to be independent of system size, underscoring the robustness of using local correlation functions as experimental signatures of crossover in current platforms \cite{Lienhard2018,Guardado2018,4}.

\subsection{Universality of local correlations}

Our analytical approach based on the ME reveals a key finding: the $n$-th order approximation $C_{R}^{(n)}$ yields identical expressions across different lattice geometries (1D chain, ring, and bifurcate), provided the system size exceeds a geometry- and order-dependent threshold. This universality emerges because the perturbative expansion, at finite order, is sensitive only to local connectivity patterns, rendering it independent of the global lattice topology.

The preceding analysis considered the contribution of a single shortest path. In realistic materials and simulators with more complex graph structures, multiple shortest paths may exist between atoms. In such cases, the net correlation function follows an \emph{algebraic law} resulting from the coherent or incoherent superposition of contributions from all relevant shortest paths \cite{Xinwang}. Crucially, while the specific algebraic form of the correlation may change, the fundamental property of universality remains valid. That is, once the system size surpasses a threshold (which itself depends on the geometry and perturbative order $n$), the functional form of $C_{R}^{(n)}$ becomes universal for a given local connectivity pattern, regardless of the global lattice structure.

As a representative case, Fig.\,\ref{fig4} summarizes the minimum system sizes required for the convergence of $C_{R=1}^{(n)}$ to its universal form: 
(i) For $n = 1$: $N \geq 2$ atoms suffice for all geometries.
(ii) For $n = 2$: $N \geq 4$ (chain), $N \geq 3$ (ring), $N \geq 6$ (bifurcate).
(iii) For $n= 3$: $N \geq 6$ (chain), $N \geq 4$ (ring), $N \geq 6$ (bifurcate).

This universality is further corroborated by our exact numerical simulations. As shown in Fig.\,\ref{fig2} ($C_{R=1}$), Fig.\,\ref{fig3}\,(a) ($C_{R=2}$), and Fig.\,\ref{fig3}\,(b) ($C_{R=3}$), the correlation functions become nearly independent of the system size for $N > 8$. The quantitative agreement between our ME results and numerical calculations confirms that the signatures of the crossover between different dynamical phases are robust and intrinsic to sufficiently large finite systems.

These findings underscore the universality of local correlations in this non-equilibrium setting. More importantly, they provide a solid foundation for employing local correlation measurements to detect and characterize dynamical quantum phase transitions in experimental Rydberg atom arrays, where global measurements may be challenging.


\section{Conclusion}

\label{sec4}

In summary, we have investigated the non-equilibrium dynamics in a Rydberg atom array following a parametric quench from a paramagnetic to an antiferromagnetic phase. The key achievement of our work is the development of a ME approach that yields analytical expressions for the local connected correlation functions, which show quantitative agreement with exact numerical simulations. We have demonstrated that these local correlations undergo a smooth but qualitative change during the crossover, serving as robust and intrinsic indicators of the underlying transition behavior in sufficiently large finite systems. This finding establishes local connected correlations as sensitive and practical probes for capturing salient features of non-equilibrium dynamics. Given that global measurements pose significant challenges in experimental settings, our proposed local detection protocol offers a directly applicable pathway for identifying the crossover phenomena in current Rydberg atom experiments.


\section*{Acknowledgment}
This work is supported by the National Natural Science Foundation of China under Grants No.12075175 and No.12575028.

\appendix
\begin{widetext}
\section{Relation between Many-Body States and Spin-Spin Correlation}

\label{SM1}

In the treatment of the Ising Hamiltonian with transverse and longitudinal fields, the connected spin-spin 
correlation function $C(\mathbf{r})$ provides a fundamental characterization of magnetic ordering. For lattice 
displacement $\mathbf{r} = (ka, la)$, this correlation is defined as:
\begin{equation}
C(\mathbf{r}) = \frac{1}{N_{\mathbf{r}}} \sum_{\langle ij \rangle_{|\mathbf{r}_i - \mathbf{r}_j| = \mathbf{r}}} \left[ \langle n_i n_j \rangle - \langle n_i \rangle \langle n_j \rangle \right]
\end{equation}
where $N_{\mathbf{r}}$ enumerates atom pairs separated by $\mathbf{r}$. This expression simplifies to the correlation of spin operators:
\begin{equation}
C(\mathbf{r}) = \frac{1}{N_{\mathbf{r}}} \sum_{\langle ij \rangle} \left[ \langle S_i^z S_j^z \rangle - \langle S_i^z \rangle \langle S_j^z \rangle \right]
\end{equation}
since $n_i = \frac{1}{2}I_i - S_i^z$ with $I_i$ the identity operator.

Next, we will demonstrate the relation between many-body states and some typical spin-spin correlations, e.g., FM, AFM, and PM correlations, by the nearest-neighbor correlation $C_{i,i+1}$. We define m-site many-body states $| \psi \rangle = \sum_{\sigma_{1},\cdots,\sigma_{i},\sigma_{i+1},\cdots,\sigma_{m}} f(\sigma_{1},\cdots,\sigma_{i},\sigma_{i+1},\cdots,\sigma_{m}) |\sigma_{1},\cdots,\sigma_{i},\sigma_{i+1},\cdots,\sigma_{m}\rangle$. Where $\sigma_{k}(k\in[1,m])=0,1$.

\noindent\textbf{(a) Fixed occupation state} (uncorrelated): 
\begin{equation}
|\psi\rangle = \sum_{\substack{\{\sigma\}\\ \sigma_i=1,\sigma_{i+1}=0}} c(\{\sigma\}) |\sigma_1,\dots,1_i,0_{i+1},\dots,\sigma_m\rangle
\end{equation}
This yields vanishing correlation:
\begin{equation}
C_{i,i+1} = \langle n_i n_{i+1} \rangle - \langle n_i \rangle \langle n_{i+1} \rangle = 0
\end{equation}

\noindent\textbf{(b) Antiferromagnetic (AFM) entangled state}:
\begin{equation}
|\psi\rangle = \sum_{\{\sigma\}} \frac{c(\{\sigma\})}{\sqrt{2}} \left( |\cdots,0_i,1_{i+1},\cdots\rangle + |\cdots,1_i,0_{i+1},\cdots\rangle \right)
\end{equation}
Characterized by negative correlation:
\begin{equation}
C_{i,i+1} = -\frac{1}{4}
\end{equation}

\noindent\textbf{(c) Ferromagnetic (FM) entangled state}:
\begin{equation}
|\psi\rangle = \sum_{\{\sigma\}} \frac{c(\{\sigma\})}{\sqrt{2}} \left( |\cdots,0_i,0_{i+1},\cdots\rangle + |\cdots,1_i,1_{i+1},\cdots\rangle \right)
\end{equation}
Exhibiting positive correlation:
\begin{equation}
C_{i,i+1} = \frac{1}{4}
\end{equation}

\noindent\textbf{(d) Paramagnetic state} (maximally disordered):
\begin{equation}
|\psi\rangle = \sum_{\{\sigma\}} \frac{c(\{\sigma\})}{2} \left( |\cdots,0_i,0_{i+1},\cdots\rangle + |\cdots,0_i,1_{i+1},\cdots\rangle + |\cdots,1_i,0_{i+1},\cdots\rangle + |\cdots,1_i,1_{i+1},\cdots\rangle \right)
\end{equation}
Showing no net correlation:
\begin{equation}
C_{i,i+1} = 0
\end{equation}


\section{Analytic Results for $C_{R=1}^{(n)}$ via ME}

\label{SM2}

In the ME of the nearest-neighbor correlation function, the order of the expansion dictates the spatial range of effective couplings. The first-order term is governed by direct nearest-neighbor interactions, while higher orders introduce contributions from progressively more distant lattice sites through the nested commutators of the Hamiltonian at different times. This structure systematically maps temporal non-locality into spatial non-locality, revealing that the strength and range of the effective interaction enhance as the expansion order increases.

\begin{equation}
	\includegraphics[scale=1.2]{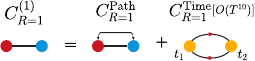}
\end{equation}
Notably, the analytic expressions exhibit a characteristic scaling structure in their denominators $\Omega^{p}\delta^{q}U^{m}T^{n}\, (p, q, m, n \in \mathbb{Z}; n = p + q + m)$,where $\Omega$ and $T$ act as universal scaling factors across all terms, while $\delta$ and $U$ vary significantly between contributions. To simplify the representation, we define a dimensionless scaling function $\mathcal{M}_{(n,p)}$ by factoring out $\Omega$ and $T$. The exact forms of each contribution for $C_{R=1}^{(1)}$ yields:

\begin{equation}
	C_{R=1}^{\rm Path}= -\frac{T^{6}\Omega^{4}}{288}\mathcal{M}_{(6,4)}
\end{equation}

\begin{equation}
	C_{R=1}^{\rm Time}[O(T^{10})]= -\frac{T^{8}\Omega^{4}}{20736}\left[ 1 + \frac{T^{2}}{288}(\delta_{f} - \delta_{i})^{2} \right](\delta_{f} - \delta_{i})^{2}\mathcal{M}_{(6,4)}.
\end{equation}
Where
\begin{equation}
	\mathcal{M}_{(6,4)}= \left[U(U-3\delta)\right].
	\notag
\end{equation}



\begin{equation}
	\includegraphics[scale=1.2]{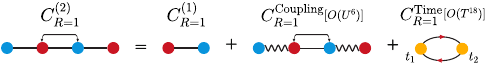}
\end{equation}
Each contribution of the correlation function in the second-order expansion can be expressed as

\begin{equation}
	\begin{split}
		C_{R=1}^{\rm Coupling}\left[O(U^{6})\right] & =  \frac{T^{8}\Omega^{4}}{11520}(\mathcal{M}_{(6,4)}\mathcal{M}_{(8,4)} + 4\Omega^{2}\mathcal{M}_{(8,6)}) \\
		& +\frac{T^{10}\Omega^{4}}{2419200}( - \mathcal{M}_{(6,4)}\mathcal{M}_{(10,4)} + \Omega^{2}\mathcal{M}_{(10,6)} + \Omega^{4}\mathcal{M}_{(10,8)})	
	\end{split}
\end{equation}

\begin{equation}
	\begin{split}
		C_{R=1}^{\rm Time}\left[O(T^{18})\right] & = \frac{T^{10}\Omega^{4}}{829440}\left[1+\frac{T^{2}}{288}(\delta_{f}-\delta_{i})^{2}\right](\delta_{f}-\delta_{i})^{2}\mathcal{M}_{(6,4)}\mathcal{M}_{(8,4)} \\
		& + \frac{T^{10}\Omega^{6}}{138240}\left[1+\frac{T^{2}}{144}(\delta_{f}-\delta_{i})^{2}+\frac{T^{4}}{62208}(\delta_{f}-\delta_{i})^{4}\right](\delta_{f}-\delta_{i})^{2}\mathcal{M}_{(8,6)} \\
	    & - \frac{T^{12}\Omega^{4}}{174182400}\left[1+\frac{T^{2}}{288}(\delta_{f}-\delta_{i})^{2}\right](\delta_{f}-\delta_{i})^{2}\mathcal{M}_{(6,4)}\mathcal{M}_{(10,4)} \\
		& + \frac{T^{12}\Omega^{6}}{116121600}\left[1+\frac{T^{2}}{144}(\delta_{f}-\delta_{i})^{2}+\frac{T^{4}}{62208}(\delta_{f}-\delta_{i})^{4}\right](\delta_{f}-\delta_{i})^{2}\mathcal{M}_{(10,6)} \\
		& + \frac{T^{12}\Omega^{8}}{87091200}\left[1+\frac{T^{2}}{96}(\delta_{f}-\delta_{i})^{2}+\frac{T^{4}}{20736}(\delta_{f}-\delta_{i})^{4}+\frac{T^{6}}{11943936}(\delta_{f}-\delta_{i})^{6}\right](\delta_{f}-\delta_{i})^{2}\mathcal{M}_{(10,8)}.
	\end{split}
\end{equation}
Where
\begin{equation}
	\begin{aligned}
	\mathcal{M}_{(8,4)} & = \left[(U^{2}-3U\delta+6\delta^{2})\right] \\
	\mathcal{M}_{(8,6)} & = \left[U(U-9\delta)\right] \\
	\mathcal{M}_{(10,4)} & = \left[3U^{4}-18U^{3}\delta+55U^{2}\delta^{2}-84U\delta^{3}+85\delta^{4}\right] \\
	\mathcal{M}_{(10,6)} & = \left[U(275U^{3}-582U^{2}\delta-123U\delta^{2}+1230\delta^{3})\right]. \\
	\mathcal{M}_{(10,8)} & = \left[U(337U+975\delta)\right].
	\notag
    \end{aligned}
\end{equation}

\begin{equation}
	\includegraphics[scale=1.2]{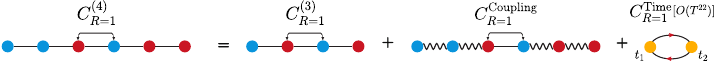}
\end{equation}
Each contribution of the correlation function in the 3rd-order expansion can be expressed as
\begin{equation}
	C_{R=1}^{\rm Coupling} = -\frac{T^{10}\Omega^{4}}{87091200}( - \mathcal{M}_{(6,4)}\mathcal{M}_{(12,4)} + \Omega^{2}\mathcal{M}_{(12,6)} + \Omega^{4}\mathcal{M}_{(12,8)} + 2\Omega^{6}\mathcal{M}_{(12,10)})
\end{equation}

\begin{equation}
	\begin{split}
		C_{R=1}^{\rm Time}\left[O(T^{22})\right] & = \frac{T^{14}\Omega^{4}}{6270566400}\left[1+\frac{T^{2}}{288}(\delta_{f}-\delta_{i})^{2}\right](\delta_{f}-\delta_{i})^{2}\mathcal{M}_{(6,4)}\mathcal{M}_{(12,4)} \\
		& - \frac{T^{14}\Omega^{6}}{4180377600}\left[1+\frac{T^{2}}{144}(\delta_{f}-\delta_{i})^{2}+\frac{T^{4}}{62208}(\delta_{f}-\delta_{i})^{4}\right](\delta_{f}-\delta_{i})^{2}\mathcal{M}_{(12,6)} \\
		& - \frac{T^{14}\Omega^{8}}{3135283200}\left[1+\frac{T^{2}}{96}(\delta_{f}-\delta_{i})^{2}+\frac{T^{4}}{20736}(\delta_{f}-\delta_{i})^{4}+\frac{T^{6}}{11943936}(\delta_{f}-\delta_{i})^{6}\right](\delta_{f}-\delta_{i})^{2}\mathcal{M}_{(12,8)} \\
		& - \frac{T^{14}\Omega^{10}}{1254113280}\left[1+\frac{T^{2}}{72}(\delta_{f}-\delta_{i})^{2}+\frac{T^{4}}{10368}(\delta_{f}-\delta_{i})^{4}+\frac{T^{6}}{2985984}(\delta_{f}-\delta_{i})^{6}+\frac{T^{8}}{2149908480}(\delta_{f}-\delta_{i})^{8}\right] \\
		& (\delta_{f}-\delta_{i})^{2}\mathcal{M}_{(12,10)}.
	\end{split}
\end{equation}
Where
\begin{equation}
	\begin{aligned}
	\mathcal{M}_{(12,4)} & = \left[U^{6}-9U^{5}\delta+40U^{4}\delta^{2}-105U^{3}\delta^{3}+178U^{2}\delta^{4}-183U\delta+124\delta^{6}\right] \\
	\mathcal{M}_{(12,6)} & = \left[U(663U^{5}-3265U^{4}\delta+6740U^{3}\delta^{2}-6033U^{2}\delta^{3}+57U\delta^{4}+3438\delta^{5})\right] \\
	\mathcal{M}_{(12,8)} & = \left[U(3946U^{3}-10905U^{2}\delta+8271U\delta^{2}+5760\delta^{3})\right] \\
	\mathcal{M}_{(12,10)} & = \left[U(1597U+1347\delta)\right].
	\notag
    \end{aligned}
\end{equation}


\section{Analytic Results for $C_{R=2,3}^{(n)}$ via ME}

\begin{equation}
	\includegraphics[scale=1.2]{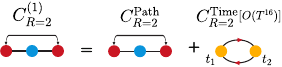}
\end{equation}
Each contribution of the correlation function $C_{R=2}^{(1)}$ can be expressed as
\begin{equation}
	C_{R=2}^{\rm Path} = \frac{T^{10}\Omega^{6}}{2419200}\mathcal{M}_{(10,6)}
\end{equation}

\begin{equation}
	\begin{split}
		C_{R=2}^{\rm Time}\left[O(T^{16})\right] = \frac{T^{12}\Omega^{4}}{116121600}\left[1+\frac{T^{2}}{144}(\delta_{f}-\delta_{i})^{2}+\frac{T^{4}}{62208}(\delta_{f}-\delta_{i})^{4}\right](\delta_{f}-\delta_{i})^{2}\mathcal{M}_{(10,6)}.
	\end{split}
\end{equation}
Where
\begin{equation}
	\begin{aligned}
	\mathcal{M}_{(10,6)} = \left[U^{2}(11U-25\delta)(7U-15\delta)\right]
	\notag
    \end{aligned}
\end{equation}



\begin{equation}
	\includegraphics[scale=1.2]{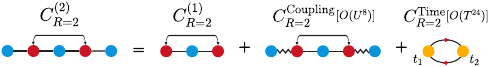}
\end{equation}
Each contribution of the correlation function in $C_{R=2}^{(2)}$ can be expressed as
\begin{equation}
    \begin{split}
	C_{R=2}^{\rm Coupling}\left[O(U^{8})\right] & = -\frac{T^{12}\Omega^{6}}{7257600}(\mathcal{M}_{(12,6)} + 4\Omega^{2}\mathcal{M}_{(12,8)}) \\
	& + \frac{T^{14}\Omega^{6}}{2235340800}(12\mathcal{M}_{(14,6)} - \Omega^{2}\mathcal{M}_{(14,8)} + 6\Omega^{4}\mathcal{M}_{(14,10)})
	\end{split}
\end{equation}

\begin{equation}
	\begin{split}
		C_{R=2}^{\rm Time}\left[O(T^{24})\right] & =-\frac{T^{14}\Omega^{8}}{1393459200}\left[1+\frac{T^{2}}{144}(\delta_{f}-\delta_{i})^{2}+\frac{T^{4}}{62208}(\delta_{f}-\delta_{i})^{4}\right](\delta_{f}-\delta_{i})^{2}\mathcal{M}_{(12,6)} \\
		& - \frac{T^{14}\Omega^{8}}{261273600}\left[1+\frac{T^{2}}{96}(\delta_{f}-\delta_{i})^{2}+\frac{T^{4}}{20736}(\delta_{f}-\delta_{i})^{4}+\frac{T^{6}}{11943936}(\delta_{f}-\delta_{i})^{6}\right](\delta_{f}-\delta_{i})^{2}\mathcal{M}_{(12,8)} \\
		& + \frac{T^{14}\Omega^{6}}{107296358400}\left[1+\frac{T^{2}}{144}(\delta_{f}-\delta_{i})^{2}+\frac{T^{4}}{62208}(\delta_{f}-\delta_{i})^{4}\right](\delta_{f}-\delta_{i})^{2}\mathcal{M}_{(14,6)} \\
		& - \frac{T^{14}\Omega^{8}}{965667225600}\left[1+\frac{T^{2}}{96}(\delta_{f}-\delta_{i})^{2}+\frac{T^{4}}{20736}(\delta_{f}-\delta_{i})^{4}+\frac{T^{6}}{11943936}(\delta_{f}-\delta_{i})^{6}\right](\delta_{f}-\delta_{i})^{2}\mathcal{M}_{(14,8)} \\
		& + \frac{T^{14}\Omega^{10}}{128755630080}\left[1+\frac{T^{2}}{72}(\delta_{f}-\delta_{i})^{2}+\frac{T^{4}}{10368}(\delta_{f}-\delta_{i})^{4}+\frac{T^{6}}{2985984}(\delta_{f}-\delta_{i})^{6} \right.\\
		&\left. + \frac{T^{8}}{2149908480}(\delta_{f}-\delta_{i})^{8}\right](\delta_{f}-\delta_{i})^{2}\mathcal{M}_{(14,10)}.
	\end{split}
\end{equation}
Where
\begin{equation}
	\begin{aligned}
	\mathcal{M}_{(12,6)} & = \left[U^{2}(52U^{4}-368U^{3}\delta+1037U^{2}\delta^{2}-1432U\delta^{3}+831\delta^{4})\right] \\
 	\mathcal{M}_{(12,8)} & = \left[U^{2}(U^{2}-143U\delta+318\delta^{2})\right] \\
	\mathcal{M}_{(14,6)} & = \left[U^{2}(121U^{6}-1170U^{5}\delta+4952U^{4}\delta^{2}-11862U^{3}\delta^{3}+17112U^{2}\delta^{4}-14322U\delta^{5}+5565\delta^{6})\right] \\
	\mathcal{M}_{(14,8)} & = \left[U^{2}(42405U^{4}-153388U^{3}\delta+102973U^{2}\delta^{2}+165984U\delta^{3}-222390\delta^{4})\right] \\
	\mathcal{M}_{(14,10)} & = \left[U^{2}(-6347U^{2}+7868U\delta+25935\delta^{2})\right].
	\notag
    \end{aligned}
\end{equation}

\begin{equation}
	\includegraphics[scale=1.2]{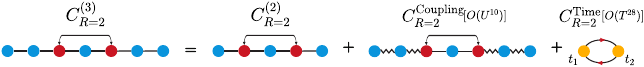}
\end{equation}
Each contribution of the correlation function in $C_{R=2}^{(3)}$ can be expressed as
\begin{equation}
    \begin{split}
	C_{R=2}^{\rm Coupling}\left[O(U^{10})\right] & =- \frac{T^{16}\Omega^{6}}{41845579776000}(3\mathcal{M}_{(16,6)} - \Omega^{2}\mathcal{M}_{(16,8)}- 2\Omega^{4}\mathcal{M}_{(16,10)}+ 48\Omega^{6}\mathcal{M}_{(16,12)})
	\end{split}
\end{equation}

\begin{equation}
	\begin{split}
		C_{R=2}^{\rm Time}\left[O(T^{28})\right] & =- \frac{T^{18}\Omega^{6}}{669529276416000}\left[1+\frac{T^{2}}{144}(\delta_{f}-\delta_{i})^{2}+\frac{T^{4}}{62208}(\delta_{f}-\delta_{i})^{4}\right](\delta_{f}-\delta_{i})^{2}\mathcal{M}_{(16,6)} \\
		& + \frac{T^{18}\Omega^{8}}{1506440871936000}\left[1+\frac{T^{2}}{96}(\delta_{f}-\delta_{i})^{2}+\frac{T^{4}}{20736}(\delta_{f}-\delta_{i})^{4}+\frac{T^{6}}{11943936}(\delta_{f}-\delta_{i})^{6}\right](\delta_{f}-\delta_{i})^{2}\mathcal{M}_{(16,8)} \\
		& + \frac{T^{18}\Omega^{10}}{602576348774400}\left[1+\frac{T^{2}}{72}(\delta_{f}-\delta_{i})^{2}+\frac{T^{4}}{10368}(\delta_{f}-\delta_{i})^{4}+\frac{T^{6}}{2985984}(\delta_{f}-\delta_{i})^{6} \right.\\
		&\left. + \frac{T^{8}}{2149908480}(\delta_{f}-\delta_{i})^{8}\right](\delta_{f}-\delta_{i})^{2}\mathcal{M}_{(16,10)}\\
		& - \frac{T^{18}\Omega^{12}}{20,922,789,888,000}\left[1+\frac{5T^{2}}{288}(\delta_{f}-\delta_{i})^{2}+\frac{5T^{4}}{31104}(\delta_{f}-\delta_{i})^{4}+\frac{5T^{6}}{5971968}(\delta_{f}-\delta_{i})^{6} \right.\\
		&\left. + \frac{T^{8}}{429981696}(\delta_{f}-\delta_{i})^{8}+ \frac{T^{10}}{7772904012465121001472000}(\delta_{f}-\delta_{i})^{10}\right](\delta_{f}-\delta_{i})^{2}\mathcal{M}_{(16,12)}.
	\end{split}
\end{equation}
Where
\begin{equation}
	\begin{aligned}
	\mathcal{M}_{(16,6)} & = \left[U^{2}(15535U^{8} - 189784U^{7} \delta + 1055262 U^{6} \delta^{2} - 3502012 U^{5} \delta^{3} + 7618415U^{4} \delta^{4} - 11190036U^{3}\delta^{5}  \right.\\
	&\left. +10925798U^{2}\delta^{6}- 6566960 U \delta^{7} + 1900758 \delta^{8})\right] \\
 	\mathcal{M}_{(16,8)} & = \left[U^{2}(6549829 U^{6} - 41061284 U^{5} \delta+ 103529220 U^{4} \delta^{2} - 127026748 U^{3} \delta^{3} + 56056843 U^{2} \delta^{4} + 28798560 U \delta^{5} \right.\\
	&\left.- 31662036 \delta^{6})\right] \\
	\mathcal{M}_{(16,10)} & = \left[U^{2}(8268429 U^{4} - 43863550 U^{3} \delta + 63939920 U^{2} \delta^{2} - 19150542 U \delta^{3} - 23108625 \delta^{4})\right] \\
	\mathcal{M}_{(16,12)} & = \left[U^{2}(-190060 U^{2} + 649117 U \delta + 422031 \delta^{2})\right].
	\notag
    \end{aligned}
\end{equation}

\begin{equation}
	\includegraphics[scale=1.2]{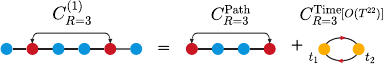}
\end{equation}
Each contribution of the correlation function $C_{R=3}^{(1)}$ can be expressed as
\begin{equation}
	C_{R=3}^{\rm Path} = \frac{T^{14}\Omega^{8}}{13412044800}\mathcal{M}_{(14,8)}'
\end{equation}

\begin{equation}
	\begin{split}
		C_{R=3}^{\rm Time}\left[O(T^{22})\right] = \frac{T^{14}\Omega^{8}}{482833612800}\left[1+\frac{T^{2}}{96}(\delta_{f}-\delta_{i})^{2}+\frac{T^{4}}{20,736}(\delta_{f}-\delta_{i})^{4}+\frac{T^{6}}{11,943,936}(\delta_{f}-\delta_{i})^{6}\right](\delta_{f}-\delta_{i})^{2}\mathcal{M}_{(14,8)}'.
	\end{split}
\end{equation}
Where
\begin{equation}
	\begin{aligned}
	\mathcal{M}_{(14,8)}' = \left[U^{3}(4279 U^{3} - 24766 U^{2} \delta + 46725 U \delta^{2} - 28350 \delta^{3})\right]
	\notag
    \end{aligned}
\end{equation}

\begin{equation}
	\includegraphics[scale=1.2]{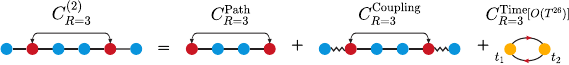}
\end{equation}
Each contribution of the correlation function in $C_{R=3}^{(2)}$ can be expressed as
\begin{equation}
    \begin{split}
	C_{R=3}^{\rm Coupling} & =- \frac{T^{16}\Omega^{8}}{10461394944000}(-\mathcal{M}_{(16,8)}' +4 \Omega^{2}\mathcal{M}_{(16,10)}')
	\end{split}
\end{equation}

\begin{equation}
	\begin{split}
		C_{R=3}^{\rm Time}\left[O(T^{28})\right] & = \frac{T^{18}\Omega^{8}}{376610217984000}\left[1+\frac{T^{2}}{96}(\delta_{f}-\delta_{i})^{2}+\frac{T^{4}}{20736}(\delta_{f}-\delta_{i})^{4}+\frac{T^{6}}{11943936}(\delta_{f}-\delta_{i})^{6}\right](\delta_{f}-\delta_{i})^{2}\mathcal{M}_{(16,8)}' \\
		& - \frac{T^{18}\Omega^{10}}{75322043596800}\left[1+\frac{T^{2}}{72}(\delta_{f}-\delta_{i})^{2}+\frac{T^{4}}{10368}(\delta_{f}-\delta_{i})^{4}+\frac{T^{6}}{2985984}(\delta_{f}-\delta_{i})^{6}\right.\\
		&\left. +\frac{T^{8}}{2149908480}(\delta_{f}-\delta_{i})^{8}\right](\delta_{f}-\delta_{i})^{2}\mathcal{M}_{(16,10)}'.
	\end{split}
\end{equation}
Where
\begin{equation}
	\begin{aligned}
	\mathcal{M}_{(16,8)}' & = \left[U^{3}(290784 U^{5} - 2409538 U^{4} \delta + 8131641 U^{3} \delta^{2} - 14177796 U^{2} \delta^{3} + 12893811 U \delta^{4} - 4857030 \delta^{5})\right] \\
 	\mathcal{M}_{(16,10)}' & = \left[U^{3}(37843 U^{3} + 390165 U^{2} \delta - 1790793 U \delta^{2} + 1614465 \delta^{3})\right].
	\notag
    \end{aligned}
\end{equation}

	\section{Adiabatic Condition}

	\label{SM4}

	We provides a rigorous mathematical derivation and verification of the quantum adiabatic condition using the ME framework. We demonstrate how the standard adiabatic condition naturally emerges from requiring the suppression of non-adiabatic effects in the ME. The quantum adiabatic theorem states that a physical system remains in its instantaneous eigenstate if a given perturbation is applied slowly enough and if there is a gap between the eigenvalue and the rest of the Hamiltonian's spectrum. The quantitative condition for adiabatic evolution is given by:

	\begin{equation}
		\frac{\hbar| \langle m(t) | \frac{d\hat{H}}{dt} | n(t) \rangle |}{|E_m(t) - E_n(t)|^{2}} \ll 1
		\label{eq:standard_condition}
	\end{equation}
	where $\hat{H}(t)$ is the time-dependent Hamiltonian, $|n(t)\rangle$ and $|m(t)\rangle$ are instantaneous eigenstates with energies $E_n(t)$ and $E_m(t)$ respectively.
	The time evolution operator $\hat{U}(t)$ can be expressed exactly using the ME:

	\begin{equation}
	\hat{U}(t) = \exp\left(\hat{\Omega}(t)\right), \quad \hat{\Omega}(t) = \sum_{k=1}^\infty \hat{\Omega}_k(t)
	\label{eq:magnus_expansion}
	\end{equation}
	
	The first few terms are:
	
	\begin{align}
	\hat{\Omega}_1(t) &= -\frac{i}{\hbar} \int_0^t dt_1 \hat{H}(t_1) \label{eq:omega1} \\
	\hat{\Omega}_2(t) &= -\frac{1}{2\hbar^2} \int_0^t dt_1 \int_0^{t_1} dt_2 [\hat{H}(t_1), \hat{H}(t_2)] \label{eq:omega2} \\
	\hat{\Omega}_3(t) &= \frac{i}{6\hbar^3} \int_0^t dt_1 \int_0^{t_1} dt_2 \int_0^{t_2} dt_3 \left([\hat{H}(t_1), [\hat{H}(t_2), \hat{H}(t_3)]] + [\hat{H}(t_3), [\hat{H}(t_2), \hat{H}(t_1)]]\right) \label{eq:omega3}
	\end{align}
	
	Each term in the ME has a clear physical interpretation:
    (i) $\hat{\Omega}_1(t)$: Represents evolution under the average Hamiltonian, corresponding to ideal adiabatic evolution.
	(ii) $\hat{\Omega}_2(t)$: First-order correction due to non-commutativity of the Hamiltonian at different times, representing the lowest-order non-adiabatic effect.
	(iii) $\hat{\Omega}_3(t)$ and higher terms: Higher-order non-adiabatic corrections.
    
	Adiabatic evolution requires that non-adiabatic effects are negligible, which in the ME framework translates to:

	\begin{equation}
	\|\hat{\Omega}_2(t) + \hat{\Omega}_3(t) + \cdots\| \ll \|\hat{\Omega}_1(t)\|
	\label{eq:magnus_condition}
	\end{equation}
	
	Specifically, for the lowest-order correction:
	
	\begin{equation}
	\|\hat{\Omega}_2(t)\| \ll \|\hat{\Omega}_1(t)\|
	\label{eq:omega2_condition}
	\end{equation}

	Assuming a total evolution time $T$ and a characteristic energy scale $E_0$ of the Hamiltonian:

	\begin{equation}
	\|\hat{\Omega}_1(t)\| \sim \frac{E_0 T}{\hbar}
	\label{eq:omega1_magnitude}
	\end{equation}

	In typical scenarios, $\|\hat{\Omega}_1(t)\| \sim 1$ (order unity phase accumulation).

	We begin with the exact expression:

	\begin{equation}
	\hat{\Omega}_2(t) = -\frac{1}{2\hbar^2} \int_0^t dt_1 \int_0^{t_1} dt_2 [\hat{H}(t_1), \hat{H}(t_2)]
	\label{eq:omega2_exact}
	\end{equation}

	Expanding $\hat{H}(t_2)$ around $t_1$:

	\begin{equation}
	\hat{H}(t_2) \approx \hat{H}(t_1) + (t_2 - t_1)\frac{d\hat{H}}{dt}\bigg|_{t_1} + \cdots
	\label{eq:h_expansion}
	\end{equation}

	Substituting into the commutator:

	\begin{equation}
	[\hat{H}(t_1), \hat{H}(t_2)] \approx (t_2 - t_1)\left[\hat{H}(t_1), \frac{d\hat{H}}{dt}\bigg|_{t_1}\right]
	\label{eq:commutator_approx}
	\end{equation}

	Thus:

	\begin{equation}
	\hat{\Omega}_2(t) \approx -\frac{1}{2\hbar^2} \int_0^t dt_1 \int_0^{t_1} dt_2 (t_2 - t_1)\left[\hat{H}(t_1), \frac{d\hat{H}}{dt}\bigg|_{t_1}\right]
	\label{eq:omega2_approx}
	\end{equation}

	\begin{equation}
		\int_0^{t_1} dt_2 (t_2 - t_1) = -\frac{1}{2}t_1^2
		\label{eq:inner_integral}
	\end{equation}
		
	Therefore:
		
	\begin{equation}
		\hat{\Omega}_2(t) \approx \frac{1}{4\hbar^2} \int_0^t dt_1 t_1^2 \left[\hat{H}(t_1), \frac{d\hat{H}}{dt}\bigg|_{t_1}\right]
		\label{eq:omega2_final_form}
	\end{equation}

	Consider transitions from initial state $|n(0)\rangle$ to $|m(0)\rangle$ ($m \neq n$):

\begin{equation}
\langle m(0) | \hat{\Omega}_2(t) | n(0) \rangle \approx \frac{1}{4\hbar^2} \int_0^t dt_1 t_1^2 \langle m(0) | \left[\hat{H}(t_1), \frac{d\hat{H}}{dt}\bigg|_{t_1}\right] | n(0) \rangle
\label{eq:matrix_element1}
\end{equation}

Under the adiabatic approximation, instantaneous eigenstates evolve slowly:

\begin{equation}
\langle m(0) | \left[\hat{H}(t_1), \frac{d\hat{H}}{dt}\bigg|_{t_1}\right] | n(0) \rangle \approx \langle m(t_1) | \left[\hat{H}(t_1), \frac{d\hat{H}}{dt}\bigg|_{t_1}\right] | n(t_1) \rangle
\label{eq:matrix_approx}
\end{equation}

Using the identity for eigenstates of $\hat{H}$:

\begin{equation}
\langle m | [\hat{H}, \hat{O}] | n \rangle = (E_m - E_n) \langle m | \hat{O} | n \rangle \quad \text{(when } \hat{H}|n\rangle = E_n|n\rangle\text{)}
\label{eq:commutator_identity}
\end{equation}

We obtain:

\begin{equation}
\langle m(t_1) | \left[\hat{H}(t_1), \frac{d\hat{H}}{dt}\bigg|_{t_1}\right] | n(t_1) \rangle = (E_m(t_1) - E_n(t_1)) \langle m(t_1) | \frac{d\hat{H}}{dt}\bigg|_{t_1} | n(t_1) \rangle
\label{eq:matrix_element2}
\end{equation}

\begin{equation}
	\langle m(0) | \hat{\Omega}_2(t) | n(0) \rangle \approx \frac{1}{4\hbar^2} \int_0^t dt_1 t_1^2 (E_m(t_1) - E_n(t_1)) \langle m(t_1) | \frac{d\hat{H}}{dt}\bigg|_{t_1} | n(t_1) \rangle
	\label{eq:final_matrix_element}
\end{equation}

Taking $t \sim T$ (total evolution time) and estimating orders of magnitude:

\begin{equation}
|\langle m(0) | \hat{\Omega}_2(T) | n(0) \rangle| \sim \frac{T^2}{\hbar^2} |E_m - E_n| \cdot \left| \langle m | \frac{d\hat{H}}{dt} | n \rangle \right|
\label{eq:order_estimate1}
\end{equation}

Since $\frac{d\hat{H}}{dt} \sim \frac{\Delta H}{T}$, where $\Delta H$ is the total variation of the Hamiltonian:

\begin{equation}
|\langle m(0) | \hat{\Omega}_2(T) | n(0) \rangle| \sim \frac{T}{\hbar^2} |E_m - E_n| \cdot |\langle m | \Delta H | n \rangle|
\label{eq:order_estimate2}
\end{equation}

Requiring $|\langle m(0) | \hat{\Omega}_2(T) | n(0) \rangle| \ll 1$ gives:

\begin{equation}
\frac{T}{\hbar^2} |E_m - E_n| \cdot |\langle m | \Delta H | n \rangle| \ll 1
\label{eq:intermediate_condition}
\end{equation}

Rearranging:

\begin{equation}
\frac{\hbar|\langle m | \Delta H | n \rangle|}{T |E_m - E_n|^{2}} \ll \frac{\hbar^{3}}{T^{2} |E_m - E_n|^{3}}
\label{eq:rearranged_condition}
\end{equation}

\begin{figure*}[htpb]
	\centering
	\includegraphics[scale=1.0]{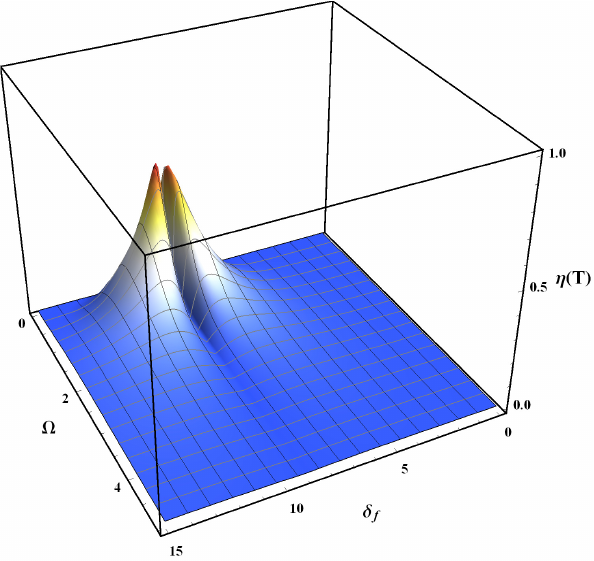}
	\caption{Adiabaticity estimation in a two-site model. The adiabatic coefficient of the system is expressed as \(\eta(T) = \frac{\hbar |\langle m | \Delta H | n \rangle|}{T |E_m - E_n|^{2}}\). With the interaction strength fixed at \(U/h = 3\,\text{MHz}\), we evaluate how the adiabatic coefficient varies with the final detuning $\delta_{f}$ and the Rabi frequency $\Omega$.}
	\label{fig5}
\end{figure*}

Since $\frac{\hbar}{T |E_m - E_n|} \ll 1$ (the core requirement of adiabatic approximation), we obtain the sufficient condition:

\begin{equation}
\frac{\hbar |\langle m | \Delta H | n \rangle|}{T |E_m - E_n|^{2}} \ll 1
\label{eq:simplified_condition}
\end{equation}

More precisely, considering the relationship between $\frac{d\hat{H}}{dt}$ and the instantaneous rate of change, we recover the standard adiabatic condition:

\begin{equation}
\frac{\hbar|\langle m(t) | \frac{d\hat{H}}{dt} | n(t) \rangle|}{|E_m(t) - E_n(t)|^{2}} \ll 1
\label{eq:final_adiabatic_condition}
\end{equation}

Based on Eq.\,\ref{eq:simplified_condition}, we analyze the adiabatic condition in a two-site model by examining the evolution of the instantaneous energy spectrum with system parameters while keeping the interaction strength fixed. For each combination of detuning $\delta$ and Rabi coupling $\Omega$, we consider the maximum energy difference between the eigenstates. As shown in Fig.\,\ref{fig5}, the adiabatic coefficient
$\eta(T) = \frac{\hbar \, \lvert\langle m | \Delta H | n \rangle\rvert}{T \, \lvert E_m - E_n \rvert^{2}}$
decreases with increasing $\Omega$ and exhibits a double-peak structure as $\delta$ varies. Physically, a larger Rabi coupling tends to open the energy gap, thereby suppressing non-adiabatic transitions. When the detuning satisfies a specific condition ($U \approx 3\delta$), singular behavior emerges, giving rise to the double-peak structure. This indicates a crossover phenomenon near this parameter regime. Consequently, in our model, the chosen range $\Omega \leq 2$ leads to a relatively large adiabatic coefficient under the implemented quench protocol, resulting in significant non-adiabatic transitions.

\section{Universality and Finite-Size Independence of $C_{R=1}$ Dynamics}
	\label{SM5}

	In our study, we employed numerical methods to investigate the dynamical evolution of the system in relation to its size by analyzing the nearest-neighbor connected correlation function. It was observed that the behavior of this correlation function remains unaffected by variations in the system size. This finding clearly demonstrates the absence of finite-size effects and indicates that the system exhibits universal scaling behavior, maintaining consistent dynamical properties regardless of the spatial extent considered.	

	\begin{figure*}[htpb]
	\centering
	\includegraphics[scale=1.0]{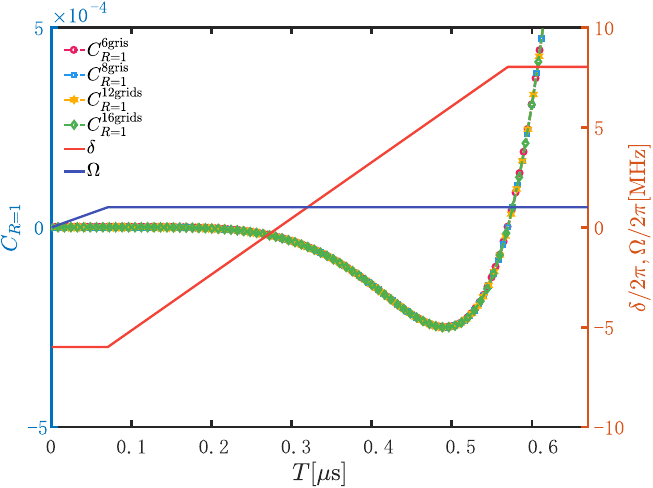}
	\caption{Comparison of the dynamical process under different lattice configurations. The nearest-neighbor connected correlation function ($C_{R=1}$) was computed numerically for various lattice sizes. Here, $C^{\rm Ngrids}_{R=1}$ represents the result of the system evolving on a lattice with N grid points.}
	\label{}
	\end{figure*}

\end{widetext}

\bibliography{DQPTs}

\end{document}